\documentclass{article}

\usepackage{arxiv}

\usepackage[utf8]{inputenc} 
\usepackage[T1]{fontenc}    
\usepackage{hyperref}       
\usepackage{url}            
\usepackage{booktabs}       
\usepackage{amsfonts}       
\usepackage{nicefrac}       
\usepackage{microtype}      
\usepackage{lipsum}
\usepackage{footmisc}       

\usepackage{xcolor}
\usepackage[square, numbers, sort&compress]{natbib}
\usepackage{graphicx}
\usepackage{array}
\usepackage{caption}
\usepackage{amsmath} 

\usepackage{dblfloatfix}

\title{Community Detection in the Hyperbolic Space}

\author{
  Matteo Bruno\thanks{These authors contributed equally to the work}\\
  IMT School \\
  Lucca 55100, Italy\\
  \And
  Sandro Ferreira Sousa\footnote[1]{}\\
  Queen Mary UoL\\
  London E1 4NS, UK\\
  \And
  Furkan Gursoy\footnote[1]{}\\
  Bo\u{g}azi\c{c}i University\\
  Istanbul 34342, Turkey\\
  \And
  Matteo Serafino\footnote[1]{}\\
  IMT School \\
  Lucca 55100, Italy\\
  \And
  Francesca V. Vianello\footnote[1]{}\\
  Imperial College London\\
  London, SW7 2AZ\\
 \And
 Ana Vrani\'{c}\footnote[1]{}\\
 Institute of Physics Belgrade,\\
 Belgrade 11080, Serbia\\
 \AND
 Mari\'{a}n Bogu\~{n}\`{a}\thanks{Corresponding author: \texttt{marian.boguna@ub.edu}}\\
 Universitat de Barcelona,\\
 Barcelona E-08028, Spain\\
}

\begin{document}
\maketitle

\begin{abstract}
Embedding a network in hyperbolic space can reveal interesting features for the network structure, especially in terms of self-similar characteristics. 
The hidden metric space, which can be thought of as the underlying structure of the network, is able to preserve some interesting features generally observed in real-world networks such as heterogeneity in the degree distribution,  high clustering coefficient, and small-world effect. Moreover, the angular distribution of the nodes in the hyperbolic plane reveals a community structure of the embedded network. It is worth noting that, while a large body of literature compares well-known community detection algorithms, there is still no consensus on what defines an ideal community partition on a network. Moreover, heuristics for communities found on networks embedded in the hyperbolic space have been investigated here for the first time. We compare the partitions found on embedded networks to the partitions obtained before the embedding step, both for a synthetic network and for two real-world networks. The second part of this paper presents the application of our pipeline to a network of retweets in the context of the Italian elections. Our results uncover a community structure reflective of the political spectrum, encouraging further research on the application of community detection heuristics to graphs mapped onto hyperbolic planes.
\end{abstract}

\keywords{Complex networks \and Hyperbolic embedding \and Community detection}

\section{Introduction}
\label{sec:intro}

Complex systems arise when the collective behavior of a given group of entities cannot be inferred from properties of the single parts. The human body is a prime example of a complex system: from genes to cells to the nervous system as a whole, we are a hierarchy of connected components. Furthermore, we are ourselves organized into a societal system which is itself complex. There is no trivial reason to why a combination of chemical elements should be able to roam the earth and make it theirs, and yet here we are. 

Naturally, scientists have preoccupied themselves about how to describe the behavior of complex systems for the best part of the last century, laying the foundations for the field now known as network theory. It has been extensively applied to reveal insights into the structure underpinning many complex systems, examples are the Internet \cite{Albert1999}, regulation patterns between metabolites \cite{Mahadevan2005} and electrical grids \cite{Pagani2013}.

Early works rooted in statistical mechanics have played a central role in characterizing fundamental properties of real-world networks: they have been shown to be scale-free \cite{Albert1999}, small-world \cite{Watts1998}, and exhibiting a non-trivial community structure \cite{fortunato12}. Furthermore, it has been shown that phase diagrams of dynamical processes and critical phenomena are drastically different when the dynamics are defined on complex networks. This observation has led to significant interest towards describing the interplay between network structure, dynamics, and the function of complex systems \cite{Dorogovtsev2008,Barrat2008}.

As the amount of data at our disposal increases exponentially, along with our knowledge on biological, social and infrastructure networks, the background of those interested in complexity has evolved. Different networks require different expertise, leading to a field that now draws on concepts from systems biology, social sciences, and engineering. However, theoretical questions still remain crucial, and there is still need for network theory to address theoretical challenges present in more traditional fields of physics and mathematics \cite{Bianconi2015}.

For instance, the development of a coherent new theory of network geometry and topology has potential practical applications, advancing the understanding of network structures. It is widely recognized that such a network theory does exist, and that it is encoded into a hidden continuous metric underlying all discrete structures of complex networks \cite{Serrano2008,Boguna2009,Krioukov2010,Papadopoulos2012,Serrano2012}. The idea is that there would be a relationship between nodes in the network such that connected nodes are closer in this hidden metric space, which is widely agreed to be hyperbolic \cite{Aste2005,Kleinberg2007,Boguna2010}. Characterizing such hyperbolicity would have crucial implications in data mining \cite{Leskovec2009,Petri2013}, the study of brain structures \cite{Gallos2011,Wedeen2012,Petri2014}, and Internet routing \cite{Kleinberg2007,Boguna2009,Boguna2010,Narayan2011}. Garc\'ia-P\'erez et al. have proposed a reliable method for faithfully embedding real-world networks into this hidden metric space, dubbed \textit{Mercator} \cite{Garcia-Perez2019}. The main advantage of this method is that it identifies hidden degrees and global parameters along angular positions and node orderings, both significant steps towards a unified mapping of complex networks.

Another important branch of network topology is that of community detection, or the problem of finding groups of nodes that are more densely connected amongst them than with the rest of the network. Of mathematical interest by itself, this problem has recently resurfaced due to the scientific community's newly found focus on, for example, social and biological networks. As previously touched upon, the real-world grounding of these networks adds a dimension to the abstract graph object, and a layer of information in the metadata. This additional information in turn sets the basis for formulating the community detection problem in terms of inference: there is a `ground truth' community structure somehow built into the graph. The focus of community detection methods then hinges on recovering this `ground truth' knowing only the graph topology \cite{Moore2017}. It is worth noting that the relationship between metadata and network structure is widely recognized to be extremely difficult to characterize \cite{Peele2017}.

Communities have applications in many fields. For instance, clustering the purchasing behavior of customers of large online retailers enables the set up of accurate recommendation systems \cite{Reddy2002}. Additionally, a lot of large-data storage solutions rely on efficient clustering of data into blocks. Clustering graphs also yields information related to single nodes, such as whether a vertex is a key player in the network it is part of. This allows for the identification of weak spots or ideal targets in a multitude of different systems \cite{fortunato12}.

In this manuscript, we aim to leverage the additional information given us by embedding networks on a hyperbolic plane to find partitions that are more accurately correlated to the underlying metadata. We first apply the \textit{Mercator} package to real-world networks describing social interactions, then partition the embedded networks into communities using several heuristics. Finally, we describe a case study where a network of retweets was constructed from a Twitter dataset collected two days before the 2018 Italian election. We show how the communities retrieved by partitioning the hyperbolic space projection are more uniform than those obtained by partitioning the original network, and relate these hyperbolic communities to the Italian political spectrum. We also describe the construction of a network out of the newly found communities, and discuss its correlation with the metadata available about the twitter users.

\section{Methodological Framework}
\label{sec:methods}
\begin{figure}
    \centering
    \includegraphics[height=0.62\textheight]{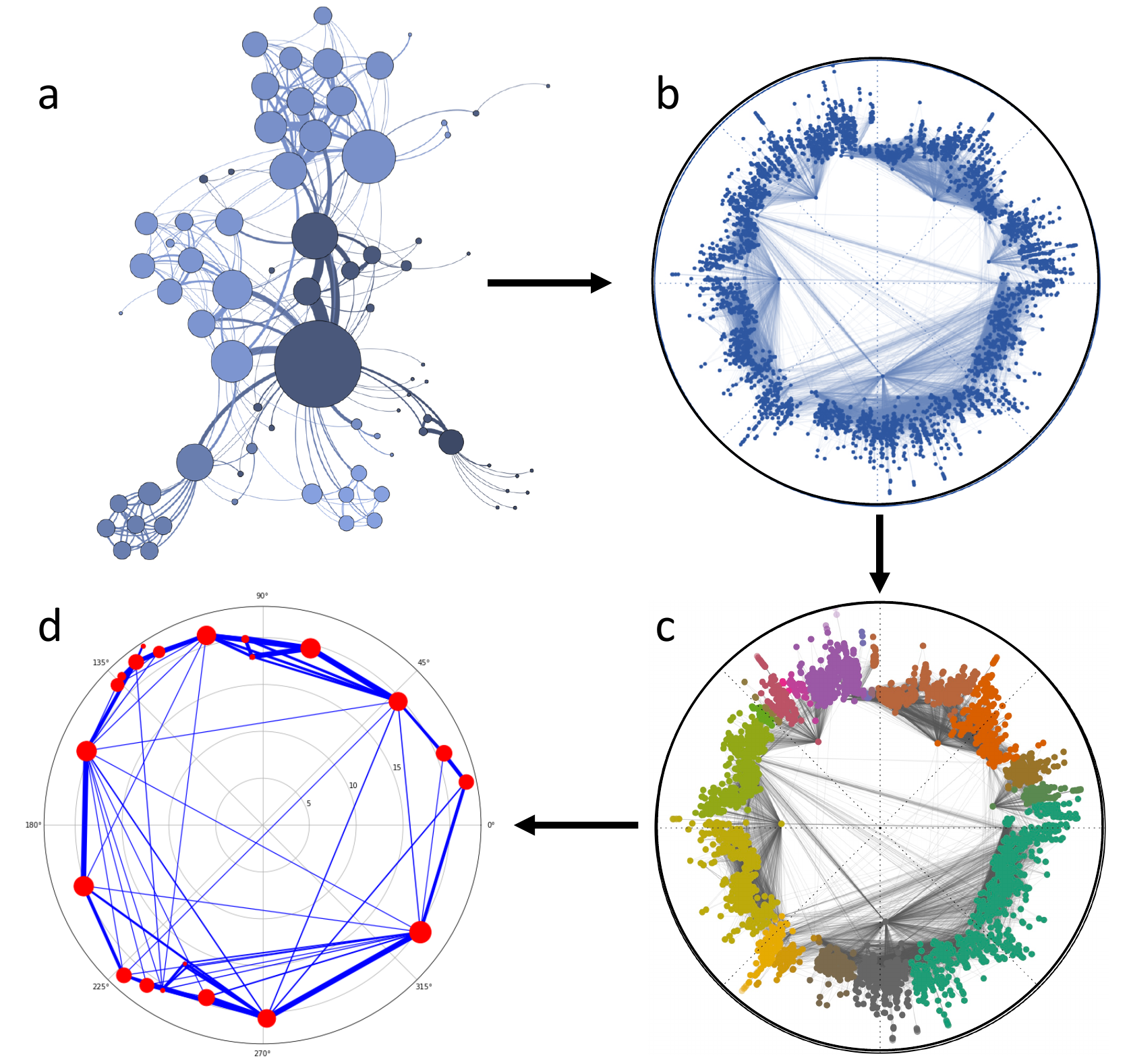}
    \caption{\textbf{Methodological pipeline.} The network is embedded on a hyperbolic plane a $\rightarrow$ b, then nodes are clustered in the new system b $\rightarrow$ c where communities are identified, and finally we construct a network from the newly formed communities c $\rightarrow$ d.}
    \label{fig:methods}
\end{figure}

We propose here a method to produce a network of communities in the hyperbolic space. It is expected that the network derived from this process will retain the system-wide information of the original graph as well as display the community structure expressed on the metadata. The process consists of three main steps, illustrated in \textbf{Figure \ref{fig:methods}}. First, we embed a network on a hyperbolic plane (\textbf{a $\rightarrow$ b}) where nodes are placed according to calculated angular position, then we cluster the new node system (\textbf{b $\rightarrow$ c}) according to a community discovery method, and finally we construct a meta network from the newly formed communities (\textbf{c $\rightarrow$ d}). The details of each step can be followed in the next sections.

\subsection{Hyperbolic embedding}
The embedding of real complex networks is not trivial. The reason is that many complex networks are not explicitly embedded in a physical space, i.e., they lack a metric structure. Generally, a metric can be defined according to the properties of the network needed to be preserved. In a recent study, Garcia-Perez et al. \cite{Garcia-Perez2019} introduced a method based on hyperbolic geometric embedding. The main hypotheses is that the architecture of real complex networks has a geometric origin defined on a metric space. Additionally, it has been shown that the $\mathbb{S}^1$ model (taking a circle as a similarity space) is the simplest among the class of geometric models \cite{Serrano2008}, yet, it preserves the system's geometric information, leading to an effective hyperbolic geometry.

Let $G(V,E)$ be a graph composed by the set of nodes $V$ and the set of vertices $E$ connecting then. The graph $G$ is connected
(otherwise, the largest component is selected) and expressed by the adjacency matrix $A$ whose elements $A_{ij}=1$ if there is a link from node $i$ to $j$, it is undirected such that $A_{ij}=A_{ji}$ and there are no self-loops ($A_{ii} = 0$). With this, the $N$ nodes of $G$ are distributed in the a circle of radius $R$ so that $\frac{N}{\rho}= 2 \pi R$. Since the choice for $\rho$ is arbitrary, it is fixed to be $\rho=1$. Then, two nodes are connected with a probability $p_{ij}$ given by:

\begin{equation} \label{eq:connect}
    p_{ij}=\frac{1}{1-\left(  \frac{d_{ij}}{\mu \kappa_i \kappa_j} \right)^\beta} 
\end{equation}

where $d_{ij}=R \Delta \theta$ is is the arc length of the circle between nodes i and j separated by an angular distance $\Delta \theta$. Parameters $\mu$ and $\beta$ control the average degree and the clustering coefficient, respectively. Even though in principle any connection probability can be used, as long as it is an integrable function, is it possible to show that the Fermi distribution defined previously ($p_{ij}$) reproduces the heterogeneity observed in empirical degree distributions together with the small world effect and the high cluster coefficient. As explained in \cite{Garcia-Perez2019}, for finite systems the values of the hidden variables $\kappa_i$ and $\theta_i$ must be evaluated numerically. The most efficient way to do so it is by combining a machine learning technique (Model-corrected Laplacian Eigenmaps) with the likelihood maximization \cite{Garcia-Perez2019}.

Since the hidden metric space formulation is independent of its $\mathbb{S}^1$ or $\mathbb{H}^2$ formulation, it is always possible to pass from one to the other. In particular the expected degree of each node $\kappa_i$ can be map to a radial coordinate as
\begin{equation}
r_i=\hat{R} - 2\ln\frac{\kappa_i}{\kappa_0}    
\end{equation}
with $\hat{R}=2\ln\frac{N}{\mu \pi \kappa_0^2}$. Using the last equation the connection probability can be written as 
\begin{equation}
p_{ij}=\frac{1}{1+e^{\frac{\beta}{2}(x_{ij}-\hat{R})}}
\end{equation}
where $x_{ij}=r_i + r_j +2\ln\frac{\Delta \theta_{ij}}{2}$ is a good approximation of the hyperbolic distance between two nodes separated by angular distance $\Delta \theta_{ij}$ and with radial coordinates $r_i$ and $r_j$.
\subsection{Community detection}
In this section, we describe two methods for community detection: critical gap method and density-based clustering. An evaluation of these approaches is discussed in \textbf{Section \ref{sec:r_d}}. The output of the embedding procedure is a tuple $(\kappa, \theta)$ for each node in the network. An interesting feature of the hyperbolic embedding is that nodes in the same angular region can be considered similar and belonging to the same cluster. However, there is no unique way to define the dimension of the angular region and no consensus on which method performs better. Besides, to assess the quality of the partition found by any community detection method, one should know the real community structure underlying the graph. These and other caveats are addressed in the Discussion (\textbf{Section \ref{sec:r_d}}).

\subsubsection{Critical gap method}
The critical gap method uses the angular distance $\Delta \theta_{ij}$ between two consecutive nodes. The main idea of this procedure is to define a critical gap $\Delta \theta_{max}$ beyond which ($\Delta \theta_{ij} > \Delta \theta_{max}$) two consecutive nodes are considered in different communities. In principle, it is possible to fix $\Delta \theta_{max}$ arbitrarily.  We modify $\Delta\theta_{max}$ from a small value so that partition contains $N$ communities to a large value so that there is only one community\footnote{The optimal way to choose $\Delta  \theta_{max}$ is to choose over all observed values of $\Delta \theta_{ij}$ of consecutive nodes}. For each of these values, we compute the communities and the modularity $Q$ \cite{PhysRevE.69.026113} according to:
\begin{equation}
Q=\frac{1}{2m} \sum_{ij} \left[  A_{ij}-\frac{k_i k_j}{2m} \right] \delta(C_i,C_j)    
\end{equation}
with $k_i$ and $k_j$ being the degrees of nodes $i$ and $j$ respectively, where the degree of a node $i$ is given by $k_i = \sum_j A_{ij}$. The variables $C_i$ and $C_j$ correspond to the respective communities and $\delta$ the delta-function. Then, we choose the partition that maximizes $Q$.

\subsubsection{Density-based clustering}
We have modified the original DBSCAN algorithm \cite{ester1996density}, whose main function is to distinguish separate dense regions, to handle (i) the angular distance rather than the euclidean distance and (ii) noise points (i.e., nodes which are not found as members of any community). For the former, we modify the distance calculation in a way that it is calculated on the circular space and hence the maximum distance between any two points is $\pi$. For the latter, for each noise point $v$, we find closest non-noise point $u$ and assign $v$ to the same community as $u$ (i.e., $c_v \gets c_u$).

The two main hyperparameters of DBSCAN are $eps$ and $min\_samples$. The readers are referred to the original paper for the details of the algorithm but basically DBSCAN defines density based on the number of nodes $min\_samples$ within a distance $eps$. To find the best hyperparameters of DBSCAN algorithm, a grid search is performed over the parameter space and modularity values are observed using the resulting partitions. For each network, the set of parameters that maximizes the modularity are used in the respective final experiments. It should also be noted that, when $min\_samples$ is equal to $1$, then DBSCAN and critical gap method usually becomes equivalent. This is due to the fact that the algorithm requires $min\_samples$ points within $eps$ distance to consider it a dense space, and if $min\_samples$ is $1$ then that is equivalent to evaluating the distance between two points based on some critical gap value which corresponds to $eps$.

\subsubsection{Other methods}

Our unreported experiments include community detection based on extreme value theory (EVT) \cite{zuev2015emergence}, k-means clustering \cite{Lloyd1982}, and agglomerative hierarchical clustering \cite{Rokach2005}. In the first method, the critical gap is estimated based on EVT rather than doing a search over the parameter space and monitoring the modularity. Our initial experiments show that it does not perform as well as our two main methods in terms of modularity. It should also be noted that the solution found by EVT-based method is already included during our parameter search in critical gap method, hence can be useful as a benchmark for modularity. On the other hand, given that modularity is not a perfect measure of real-world communities, EVT-based method might be further improved to be a useful tool for detecting real-world communities.

The community detection method which uses k-means clustering suffers from instability of the final clusters which depend on the initial selection of cluster centers. This instability was also the case with the data we have. This limitation could be addressed by sampling a very large number of initial conditions for the clustering algorithm, which was not possible under the time constraints of this workshop. Lastly, the agglomerative hierarchical clustering algorithm did not yield superior results in our limited initial experiments either. Therefore, for the scope of this work, we have excluded these two methods as well as the EVT-based method. We refer the interested reader to the Conclusion for a discussion of potential avenues for further work.

\subsection{Network of communities}
To generate the network among communities we first define an attraction potential energy between nodes, which is the logarithm of the probability of the absence of a link between them:
\begin{equation}
    V(v_i, v_j) = \ln (1 - p_{ij})
\end{equation}
where $p_{ij}$ is the probability of the presence of the link between the nodes $i$ and $j$. The potential energy between two possible sets of links will then be the logarithm of the probability of absence of all links between the sets:
\begin{equation}
    V(C, C') = \ln \left[ \prod\limits_{i \in C,\: j \in C'}(1 - p_{ij}) \right] = \sum\limits_{i \in C,\: j \in C'} \ln (1 - p_{ij}) \: .
\end{equation}
Now we can define the realized energy between sets, that will sum the contribution only of present links:
\begin{equation}
    R(C, C') = \sum\limits_{i \in C,\: j \in C'} a_{ij} \ln (1 - p_{ij}) \: .
\end{equation}

\section{Results and Discussion}
\label{sec:r_d}

\subsection{Evaluating community detection approaches}

Considering the three-step process defined previously (See \textbf{Figure \ref{fig:methods}}), we embed a series of networks in the hyperbolic space and obtain a number of communities partitioning each system. \textbf{Table \ref{tab:mod}} reports the properties of the optimal network partitions, where distinct community detection approaches are compared and benchmarked. To evaluate the partitions given by the hyperbolic embedding step, we compute their modularity for each iteration of the two community detection methods. The optimal partition for both methods is then obtained by maximizing modularity, either iterating over values of angular distance or node density. Additionally, results obtained from the Louvain-optimized partition \cite{blondel08} of the non-embedded network are reported in \textbf{Table \ref{tab:mod}}, where the distinct methods are denoted by `gap' (critical gap method), `dbscan' (density-based clustering) and `non-emb' (non-embedded network).

\begin{table}[b]
    \centering
    {\setlength{\extrarowheight}{0.3em}
    \caption{Partitions with maximum modularity across networks}
    \label{tab:mod}
    \begin{tabular}{|c||c|c|c|c|c|c|c|c|}
        \hline 
        & \multicolumn{2}{c|}{gt} & \multicolumn{2}{c|}{gap} & \multicolumn{2}{c|}{dbscan} & \multicolumn{2}{c|}{non-emb} \\ [0.15em]
        \hline
        Network & NC & $Q$ & NC & $Q$ & NC & $Q$ & NC & $Q$ \\ [0.25em]
        \hline \hline 
        Test & 7 & 0.404 & 18 & 0.554 & 18 & 0.554 & 9 & 0.592 \\[0.25em]
        \hline
        Email & 42 & 0.288 & 9 & 0.371 & 8 & 0.391 & 7 & 0.414 \\ [0.25em]
        \hline
        Facebook & \multicolumn{2}{c|}{-} & 22 & 0.746 & 19 & 0.771 & 16 & 0.835 \\[0.25em]
        \hline 
   \end{tabular}}
    \caption*{\\ \textbf{gt}: ground truth partition (see text for details); \textbf{gap}: critical gap modularity optimization; \textbf{dbscan}: density-based modularity optimization; \textbf{non-emb}: maximum modularity (Louvain algorithm); \textbf{NC}: Number of communities; \textbf{Q}: Modularity value.}
\end{table}

In addition, we report the modularity of the partition obtained by the network's `ground truth' (denoted by `gt' in \textbf{Table \ref{tab:mod}}), i.e., a specific partition tied to the process that generates each network. In fact, there is an intense debate in the literature on what should be considered the `real' community structure. For instance, cases where real-world data is anonymized will result in datasets containing numerical information on node's membership to communities, which limits interpretations beyond quantitative ID-matching approaches. Such node-ID based approaches would be of no help towards a critical evaluation of discrepancies in partitions resulting from different clustering methods. In other cases, especially in the context of social networks (i.e., links are social ties), the reported communities, often self-assigned, are not reliable and should be taken carefully as the reference configuration.

The networks used in this manuscript are as following: a synthetic test data set (with predefined community structure), an academic email collaboration network, and a network of social interactions (from Facebook). These are indicated respectively as `Test', `Email' and `Facebook' in \textbf{Table \ref{tab:mod}} and in the text hereafter. Origins and characteristics of these datasets are all detailed in \textbf{Appendix \ref{sec:app}}. The `Test' network is a synthetic network designed to have a weak community structure consisting of seven communities. On the other hand, the `Email' network's predefined community structure arises from real-world information: each edge connects two emailing agents (the nodes) at the heart of an academic institution. The nodes have identified themselves as being part of one department or the other, and the IDs of these departments have been successively anonymized. The `Facebook' dataset did not contain any metadata that could be related to non-overlapping communities, thus its `ground truth' state is unknown. 

\begin{figure}[b!]
    \centering
    \includegraphics[height=0.4\textheight, origin=c]{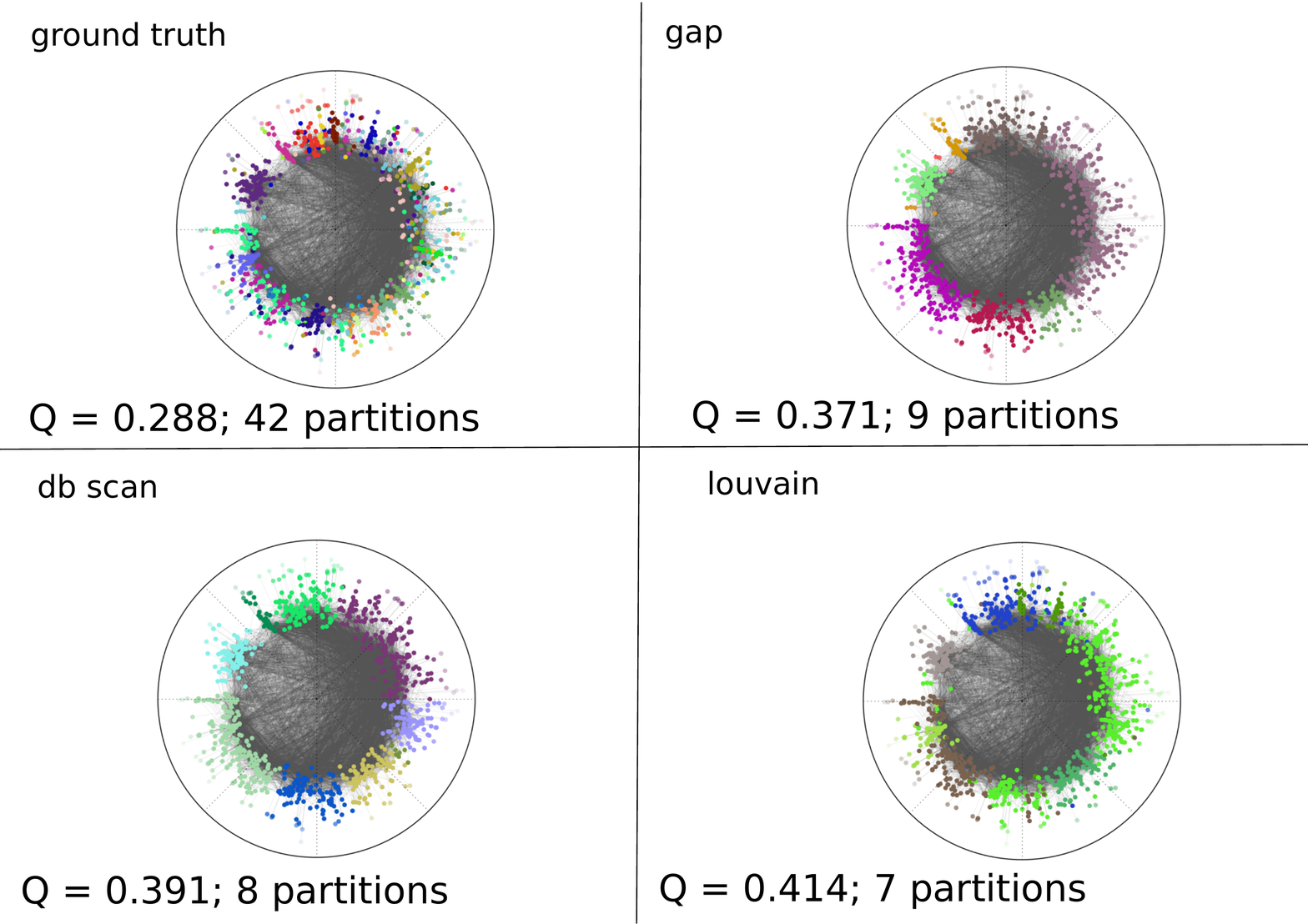}
    \caption{Email network communities. After embedding network to hyperbolic plane, we used different methods to find communities.  Note that the colors are randomly generated.}
    \label{fig:email}
\end{figure}

The projections into polar coordinates of the community partitions corresponding to \textbf{Table \ref{tab:mod}} are shown in \textbf{Figure \ref{fig:benchmark}}, in \textbf{Appendix B}. Additionally, here \textbf{Figure \ref{fig:email}} presents the subset of these projections pertinent to the `Email' network data. The modularity value ($Q$ in the figure) is the highest for the Louvain-obtained partition, as is expected. Indeed, the algorithm is designed to find the highest modularity value, and optimizes the clustering towards that. On the other hand, both critical gap method and density-based clustering aim to optimize other measures, and are therefore bound to score slightly lower in modularity. Crucially, the `ground truth' partition is radically different from the algorithmically-obtained clusterings: it has significantly lower modularity and a much higher number of communities. This is a really good indication that optimizing towards modularity does not necessarily result in the recovering of accurate metadata-inferred partitions. More importantly to us, the similarity of partition boundaries between the communities that have been obtained on the embedded graph and among the communities obtained before the embedding step confirms our original intuition: hyperbolic embedding of networks maintains the community structure.

\subsection{A case study on Italian political Twitter network}

\begin{figure}[!b]
    \centering
    \includegraphics[width=1\textwidth]{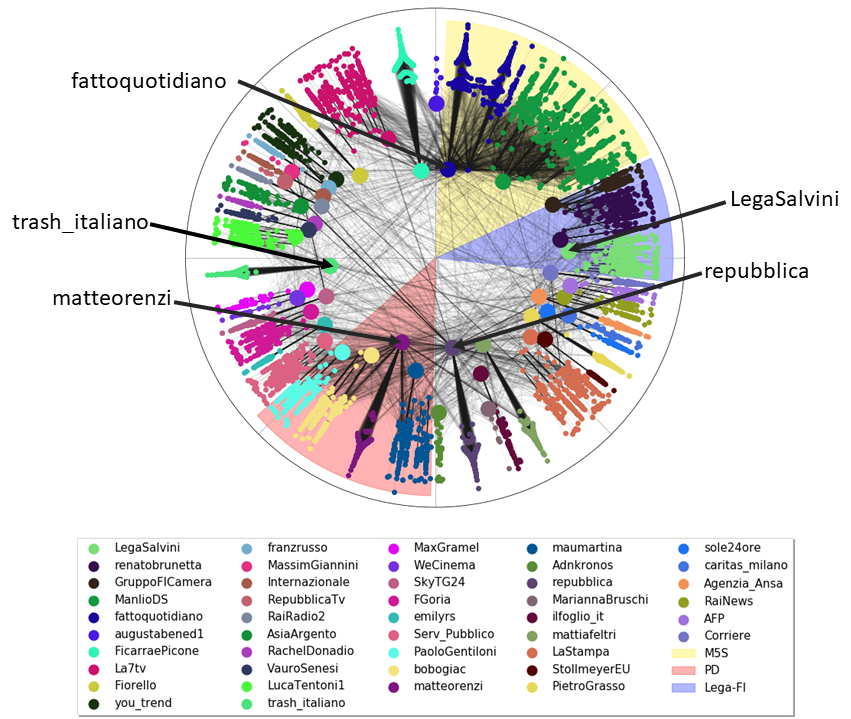}
    \caption{Twitter network embedded on a hyperbolic plane where communities are detected according to the angular region of nodes via the critical gap method and represented in different color. The legend starts with the community at zero degrees (right side), and lists the communities in anticlockwise order. The node with the highest degree in each community is represented with a larger dot and put in the legend, representing its cluster. The background colors are representing the angular sectors occupied by nodes that have a political orientation, as described in the text.}
    \label{fig:twitter_embed}
\end{figure}

We analyzed a Twitter dataset consisting of tweets from two days right before the Italian political elections in 2018, curated by Becatti et al. \cite{becatti2019extracting}. We selected retweets between users of which at least one of them was a verified user, and then linked the two users in an undirected and unweighted network. The resulting network was disconnected, so we kept the largest connected component, which comprises $\sim$90\% of nodes. This procedure yielded a network of $\sim$8000 nodes and $\sim$12000 links. We then computed the \textit{Mercator} embedding and performed community detection via the critical gap method (see \textbf{Section \ref{sec:methods}}).

Pleasingly, the so-found communities capture the Italian political scenario, creating a political compass, even in a very short two-day time span. \textbf{Figure \ref{fig:twitter_embed}} shows the radial distribution of the communities found by partitioning of the embedded network. The nodes representing the users belonging to the moderate left wing, for example, are clustered in the lower part of the embedding (shaded in red in the figure), and they are split into four communities. The node with the highest degree in each community is shown as a larger node, and within this sector of the radial plot the communities are ``represented'' by four of the main exponents of PD, the major party: \textit{Renzi, Martina, Giachetti} and \textit{Gentiloni}. It is worth noting that these politicians were at the time political competitors.

Communities of members of the moderate right wing are also neighbours in this embedded representation: we have indicated them by shading the sector in blue in \textbf{Figure \ref{fig:twitter_embed}}, and their representatives include \textit{Renato Brunetta}, and an account named after \textit{Lega Salvini}. Similarly, the independent party \textit{Movimento 5 Stelle} is located on the upper-right quadrant of the plot (shaded in yellow in the figure). The high density of this region suggests that their presence on Twitter was relevant in the analyzed days. 

Furthermore, we note the presence of a few famous comedians and satiric pages, such as \textit{trash\_italiano [sic.]} and \textit{FicarraePicone [sic.]}. Their high degrees are explained by the numerous retweets by non-verified users. Additionally the account for \textit{Repubblica}, one of the major Italian newspapers, is amongst the most central in this network.

\subsubsection{Community detection comparison}
We also computed the communities via the Louvain algorithm (shown in \textbf{Figure \ref{fig:twitter_lou}} in the appendix), finding a slightly smaller number of communities (42 against 44) but with some of the biggest communities merged. To validate this intuition, we use a measure of inclusion between two partitions \cite{Bruno_2018}:
\begin{equation}
    \operatorname{inc}(\mathcal{A}, \mathcal{B})=\frac{\sum\limits_{A\in \mathcal{A}}\max\limits_{B \in \mathcal{B}}|A\cap B|}{\sum\limits_{A\in \mathcal{A}}|A|} \: .
\end{equation}
This measure is of simple interpretation, yielding the fraction of nodes that are correctly relabeled when making a best correspondence between them. It is not symmetric so in our case the measure scores $\sim$0.76 when including our partition in the Louvain one, and $\sim$0.58 in the other case. This means that our partition is mostly a refinement of the Louvain one. While both make sense when analyzing the political divisions, it is nice to see that our subdivisions also have a meaning: for instance, the left wing party \textit{PD} is subdivided according to the four most prominent (and competing) personalities.

\subsubsection{Building the network of communities}
\begin{figure}
    \centering
    \includegraphics[width=\textwidth]{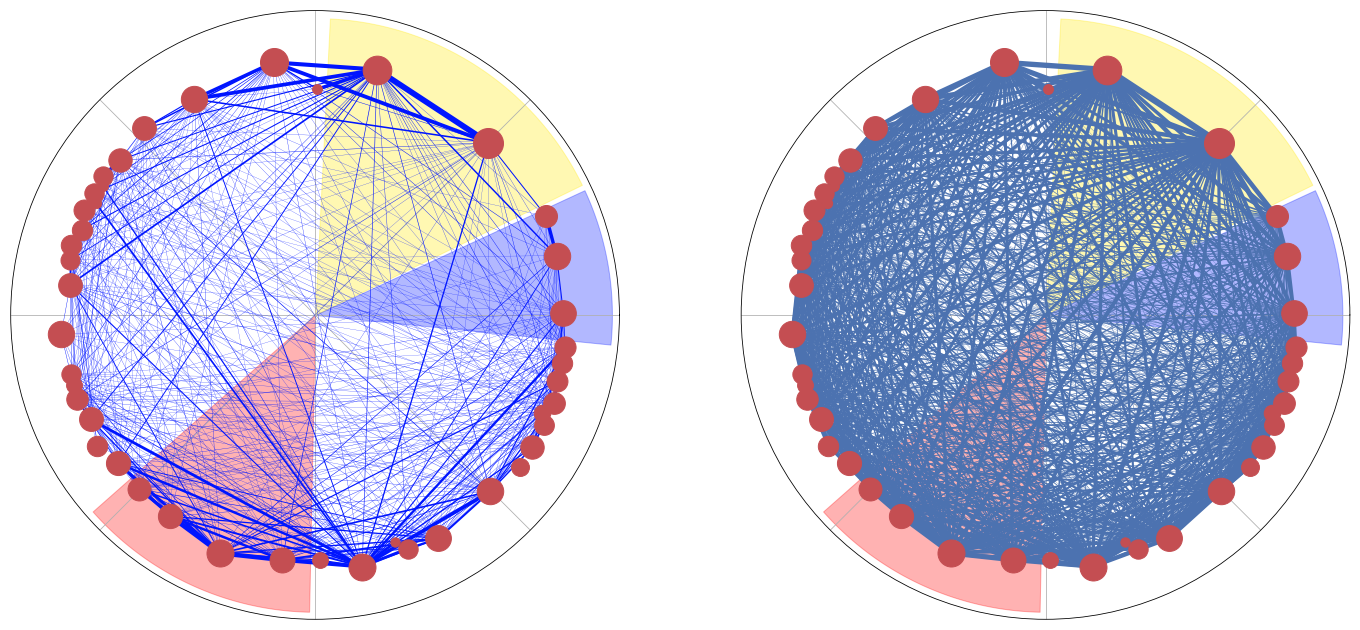}
    \caption{The network of communities generated from the Twitter data, considering the realized energies (left) and the potential energies (right). The size of the nodes is proportional to the size of the community in logarithmic scale, and their positions are the geometric centers of the nodes of the community. The width of the links represents the energy two communities exchange, while the background colors are the same as in Fig. \ref{fig:twitter_embed}. Although the realized energy network is very dense due to the density of links in the system, it gives already some nice hints: the stream of tweets between the populist party (yellow) and the right wing party (blue) is low even if in the embedded network they are close.}
    \label{fig:twitter_coarse}
\end{figure}

Following the pipeline we have detailed previously (see \textbf{Figure \ref{fig:methods}}), a network was constructed from the communities we obtained by partitioning the embedded graph, to quantify the interactions between communities and detect the structure of political forces. This network is shown in \textbf{Figure \ref{fig:twitter_coarse}}.

While it is not straightforward to interpret the results of this procedure due to the high noise in the system, it can surely be noted that even though the positions of \textit{Movimento 5 Stelle} and that of the right-wing representatives are quite near in terms of the angular distance the interaction between the respective communities is low. In the wake of the 2018 elections, many dubbed the alliance between \textit{La Lega} and \textit{Movimento 5 Stelle} to be surprising. This could be a consequence of the low number of interactions, which signify a low number of retweets between users identifying with one or the other political group. However, their coalition is supported by their position within the embedded network. This suggests a potential of geometric representation of network, and especially of hyperbolic embedding to forecast future node associations in real-world data.

\section{Conclusion}

In this work we performed several experiments using \textit{Mercator}, a hyperbolic network embedding method, on a variety of real-world networks along with a synthetic network. After projecting the networks into a latent hyperbolic space, the learned embeddings are used to find communities by utilizing different community detection methods. The findings show that the communities found on the latent space are comparable with communities found on the network using the Louvain algorithm in terms of modularity.

Although we have employed modularity as a measure of partitioning quality, modularity suffers from resolution limits and tends to favor larger communities thus may result in a relatively small number of communities, even when the communities are well defined \cite{Kumpula2007}. Confirming this, we found that the number of communities found by our community detection methods is greater than the number of communities found by the Louvain algorithm in all experiments.

It is also not straightforward to conclude that a partitioning that maximizes modularity can accurately reflect the real-world network mechanisms. This is in line with our findings where we find that aiming to maximize modularity results in greater modularity score in comparison to the ground truth. 

Among the two community detection methods, the density-based clustering approach performs as good as or slightly better than the critical gap approach in terms of modularity. Therefore, it can be concluded that the density-based clustering approach is more robust to different networks when compared to the critical gap approach, although the marginal improvement can be trivial in many cases. 

Apart from quantitative comparisons, the embeddings in the hyperbolic space where nodes are colored based on their communities depict that communities are separated quite clearly, and differences with Louvain communities exist. This case is presented on Italian political Twitter network. The Louvain algorithm finds more communities in the network, but the refinement of them that we found seems to make sense based on subdivisions and currents inside the same parties, although the data is considering only a very short time span. 

The aggregation of the communities in a new network, made using the probabilities of the Mercator embedding also gives a better insight on the structure of the political scenario of the dataset. Preliminary results also suggested the possibility for such an embedding to be able to identify hidden relationships between socio-political nodes, that were confirmed by global event. If confirmed by further work on a wider selection of real-world data, this would be an incredibly exciting application of the hyperbolic metric space for forecasting purposes. 

Overall, the findings show that the approach of detecting communities in the latent space is promising in terms of identifying and explaining the real-world communities. In accordance, future efforts could be directed towards further developing the partitioning algorithms which work on the latent space and aims to find the actual community structure rather than blindly maximizes the modularity score. As mentioned, case studies with other empirical networks can also help to evaluate the extent of generalizability of our approach into other networks.


\section{Acknowledgements}
This work is the output of the Complexity 72H Workshop, held at IMT School for Advanced Studies Lucca, 17-21 June 2019. All authors are grateful to Alberto Antonioni, Eugenio Valdano, Tiziano Squartini, Rossana Mastrandrea and IMT Lucca for giving us the opportunity to conduct this research. Website can be found at \texttt{https://complexity72h.weebly.com}

\newpage
\bibliography{geom_embedding}  

\newpage
\appendix
\section*{Appendix}
\section{Network Datasets}
\label{sec:app}
In this study, we use data extracted from the Stanford Large Network Dataset Collection (SNAP) \cite{snapnets}, the interested reader can obtain a copy from the online repository. 

\textbf{Facebook. }The Facebook Social circles dataset consisting of profile and network data from 10 ego-networks corresponds to 193 circles and 4,039 users. The 4039 nodes are defined by users and the 88234 edges by friendship relations. It was collected during a survey of ten users, who were asked to manually identify all the circles to which their friends belonged.

\textbf{Email. }The EU email communication network consists of all incoming and outgoing emails between institution members (the core) of a large European research institution. It was collected from October 2003 to May 2005 (18 months) and contains 3,038,531 emails between 287,755 different addresses. The core consists of 1005 nodes with 25,571 edges and the metadata represents the anonymized member's department (42 distinct departments). A directed edge between nodes $i$ and $j$ is created if $i$ send at least one message to $j$, however, the undirected version of this network is considered. 

\textbf{Test. }The synthetic network was generated with fixed parameters so that a specific number of communities could be obtained. It is given by defining the number of nodes $N$, the parameters $\beta$, $\langle k \rangle$, and $\gamma$ being respectively 1000, 3, 10 and 2.5. The connectivity probability $p$ is calculated according to \textbf{Eq.} \eqref{eq:connect}. Then, communities are obtained by distributing nodes randomly within a region of the hyperbolic space with angular distance $\theta$ assigned at random. Nodes are connected with probability $p$ and an `empty space' is placed between communities so that nodes belonging to each community are confined in a specific region. The choices for the synthetic network produced a graph with 7 partitions, ranging from larger communities of 563 nodes to smaller ones of 6 nodes. This settings provides a partition structure which is not trivial to detect, i.e, the density of points within the regions is close to one so that there is no evident signal to detect.

\newpage
\section{Supplementary Figures}

\begin{figure}[h!]
    \centering
    \includegraphics[height=0.8\textheight,origin=c]{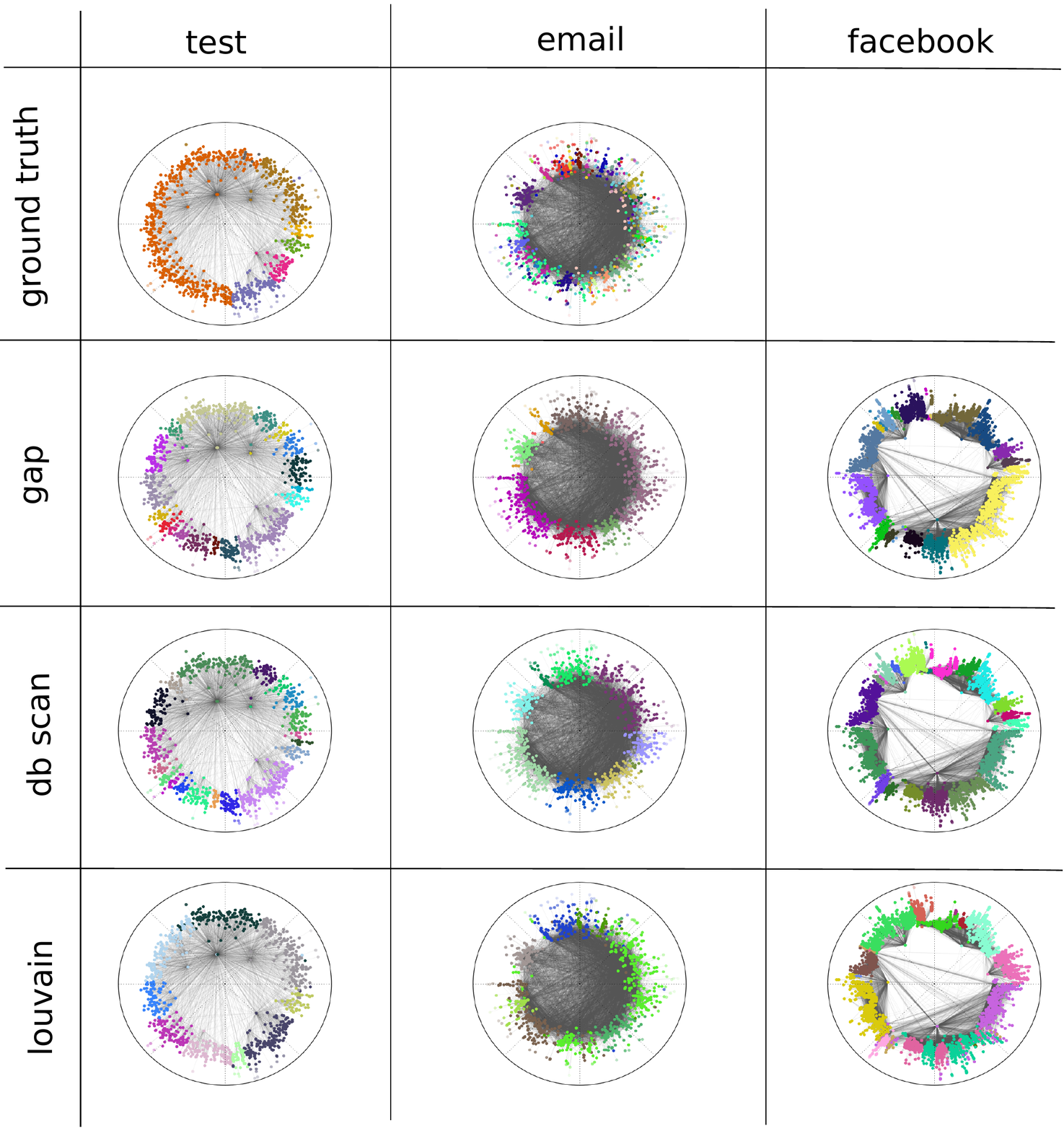}
    \caption{Facebook, email and test (synthetic) networks are embedded to a hyperbolic plane and using different methods are obtained communities. For email and test network are known ground truth partitions. For facebook network critical method gives 22 communities, density-based clustering method (dbscan) 19 of them and Louvain method only 16 partitions. Email network has 42 communities, and community detection methods obtain 9, 8 and 7 communities. Test network is created with 7 partitions, but detected numbers of partitions by different methods are 18 (gap), 18 (dbscan) and 7 (Louvain). }
    \label{fig:benchmark}
\end{figure}

\begin{figure}[h!]
    \centering
    \includegraphics[height=0.8\textheight]{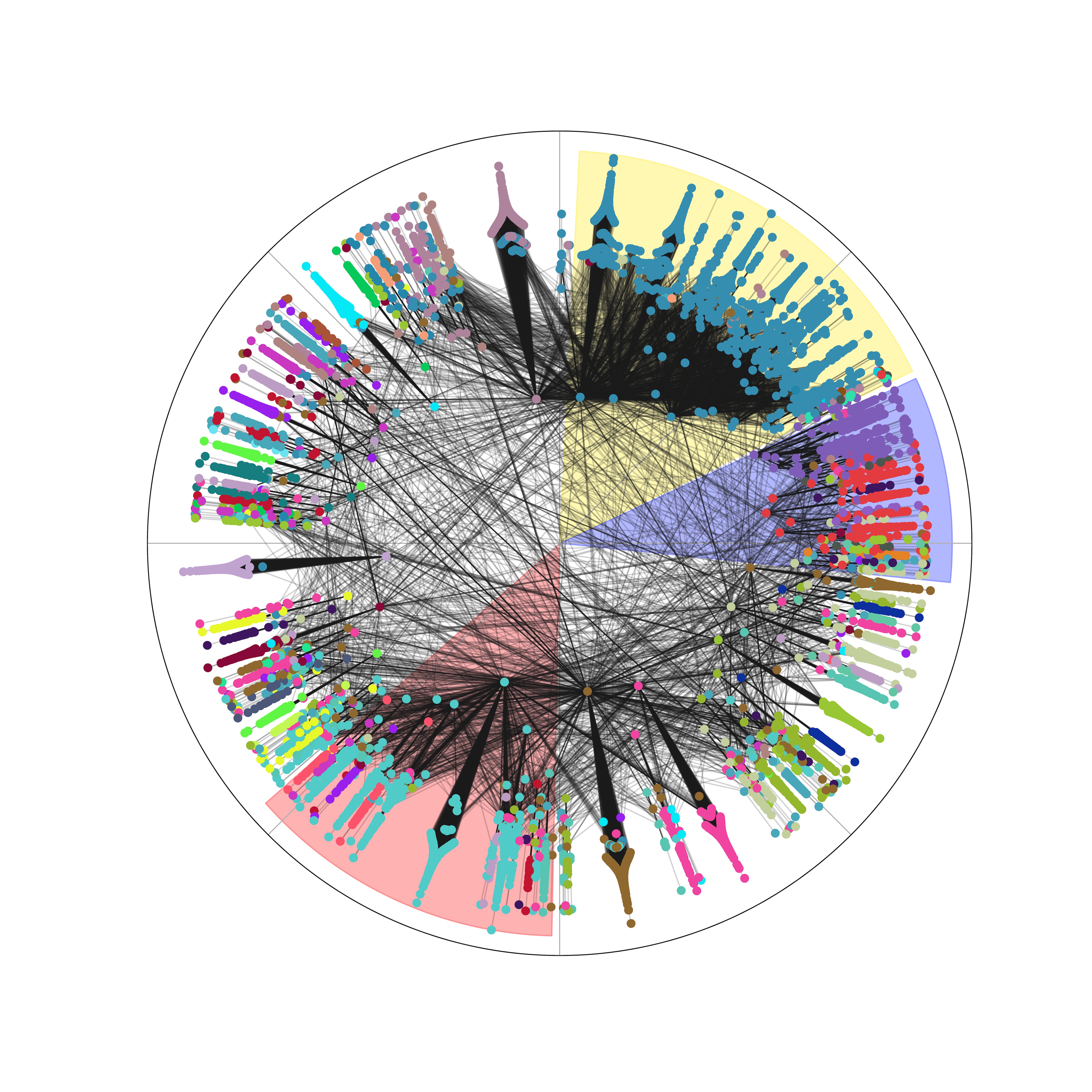}
    \caption{Communities on the Italian Political Twitter dataset, obtained by Louvain optimisation, projected onto the hyperbolic space. When compared to the underlying metadata (political factions as shown by the shaded areas), the match is noisier than when the partitioning step had been conducted after embedding the network.}
    \label{fig:twitter_lou}
\end{figure}

\end{document}